\documentclass{PoS}

\usepackage{epsfig}

\PoS{PoS(HEP2005)106}
\newcommand{\bea}{\begin{eqnarray}}
\newcommand{\eea}{\end{eqnarray}}
\newcommand{\be}{\begin{equation}}
\newcommand{\ee}{\end{equation}}

\title{On the structure of $D_{sJ}^*(2317)$ and $D_{sJ}(2460)$} 

\ShortTitle{On the structure of $D_{sJ}^*(2317)$ and $D_{sJ}(2460)$}

\author{\speaker{Pietro Colangelo}\\
        Istituto Nazionale di Fisica Nucleare,  Sezione di Bari, Italy\\ 
    E-mail: \email{pietro.colangelo@ba.infn.it}}

\author{Fulvia De Fazio\\
         Istituto Nazionale di Fisica Nucleare, Sezione di Bari, Italy\\
    E-mail: \email{fulvia.defazio@ba.infn.it}}

\author{Rossella Ferrandes\\
        Dipartimento di Fisica dell'Universit\'a di Bari and  INFN, Sezione di Bari, Italy\\
    E-mail: \email{rossella.ferrandes@ba.infn.it}}
    
\author{Altug Ozpineci\\
         INFN, Sezione di Bari, Italy and METU University, Ankara, Turkey\\
    E-mail: \email{ozpineci@ba.infn.it}}
    
\abstract{We discuss the structure of the narrow  mesons with
charm and strangeness $D_{sJ}^*(2317)$ and $D_{sJ}(2460)$,  stressing the role
 of  the radiative decay modes  in shedding light on the quark content of these states.
  We  argue  that present experimental data favour the interpretation of the two states
as ordinary $\bar c s$ mesons. We suggest
the existence of  resonances with similar properties and well predicted masses in the $\bar b  s$ sector.}

\FullConference{International Europhysics Conference on High Energy Physics\\
		 July 21st - 27th 2005\\
		 Lisboa, Portugal}

\begin{document}
\section{Introduction}

The observation of two narrow resonances with charm and strangeness,
$D^*_{sJ}(2317)$ in the  $D_s\pi^0$ channel
and $D_{sJ}(2460)$  in the $D^*_s \pi^0$ and $D_s\gamma$ channels  \cite{Aubert:2003fg},
has raised discussions about the quark  structure  of these states 
\cite{Colangelo:2004vu}.
Their identification with the scalar and axial vector $\bar c  s$ mesons  ($D_{s0}$ and $D_{s1}^\prime$)
is   natural;  in the  $m_c \to \infty$ limit such states are
expected to be degenerate in mass and to form a 
$s_\ell^P= \frac{1}{2}^+$ doublet,  with  $s_\ell$   the angular momentum of the light
degrees of freedom. In that interpretation $D_{sJ}^*(2317)$, 
 $D_{sJ}(2460)$,   $D_{s1}(2536)$
and  $D_{s2}(2573)$ are the  four  lowest lying P-wave $\bar c s$ states.
However, quark model estimates of the masses of these mesons
 generally  produce larger values  than the measured ones, implying  that the two $\bar c s$
 scalar and axial-vector   $D_{s0}$ and $D_{s1}^\prime$ states should be
heavy enough  to  decay  to $D K$ and $D^* K$.
 Moreover, the   $B \to D^*_{sJ}(2317) D$ and $B \to D_{sJ}(2460) D$ decay rates 
computed by  naive factorization  do not agree with  experiment  \cite{Datta:2003re}. 
 On this basis,  other interpretations  for $D^*_{sJ}(2317)$ and $D_{sJ}(2460)$ have been proposed,  namely  that of  molecular states \cite{close}.
Radiative transitions  probe the structure of hadrons and are 
suitable  to understand  the nature of $D_{sJ}^*(2317)$ and $D_{sJ}(2460)$  
since the rates
can be predicted by  various methods  and compared to the
experimental findings 
 \cite{Godfrey:2003kg,colangelo1}.  One method is  Light-Cone QCD sum rules 
\cite{Colangelo:2005hv}.

\section{Radiative decays by Light-Cone  QCD Sum Rules}
\begin{figure}[tbhp]
\begin{center}
\begin{tabular}{cc}
\epsfig{file=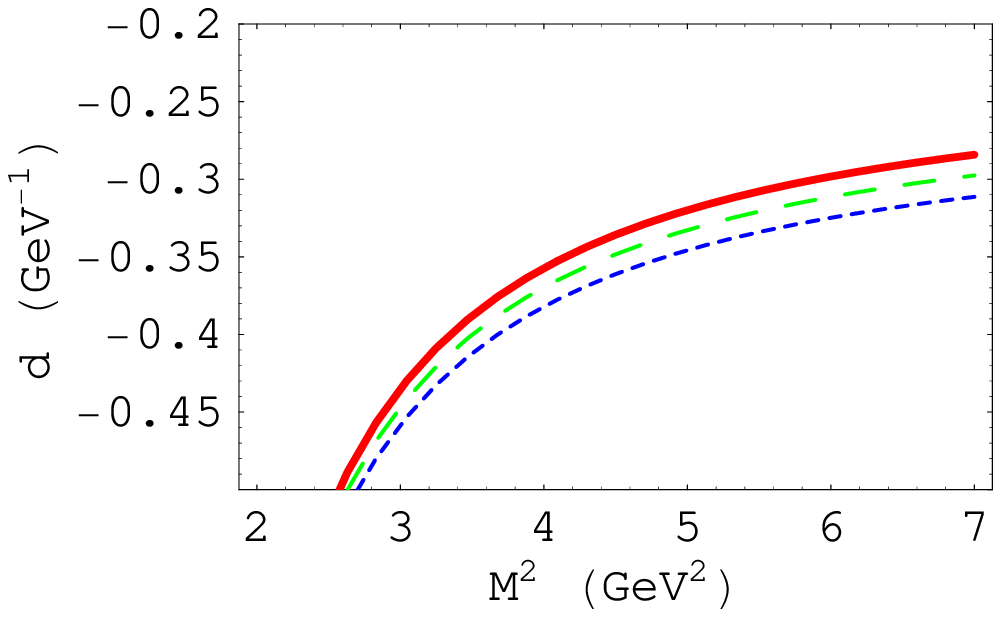, width=0.3\textwidth}&\epsfig{file=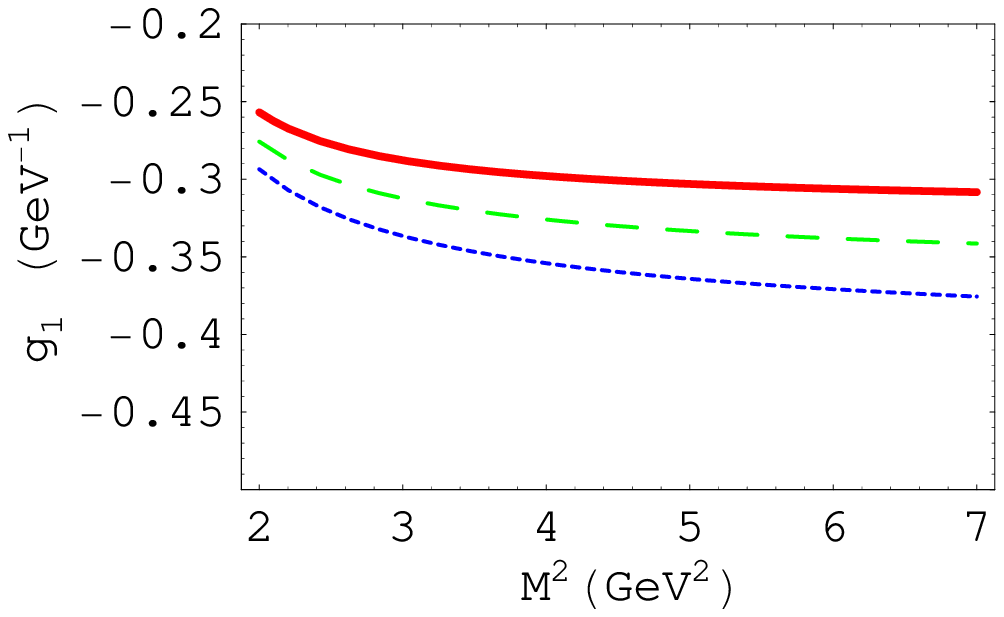, width=0.3\textwidth}\\
\epsfig{file=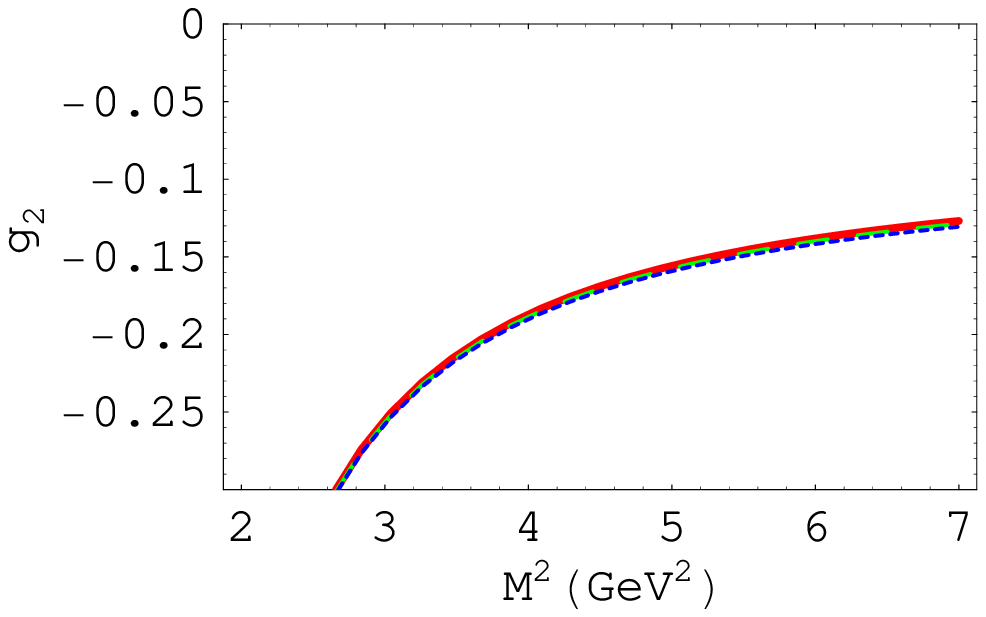, width=0.3\textwidth}& \epsfig{file=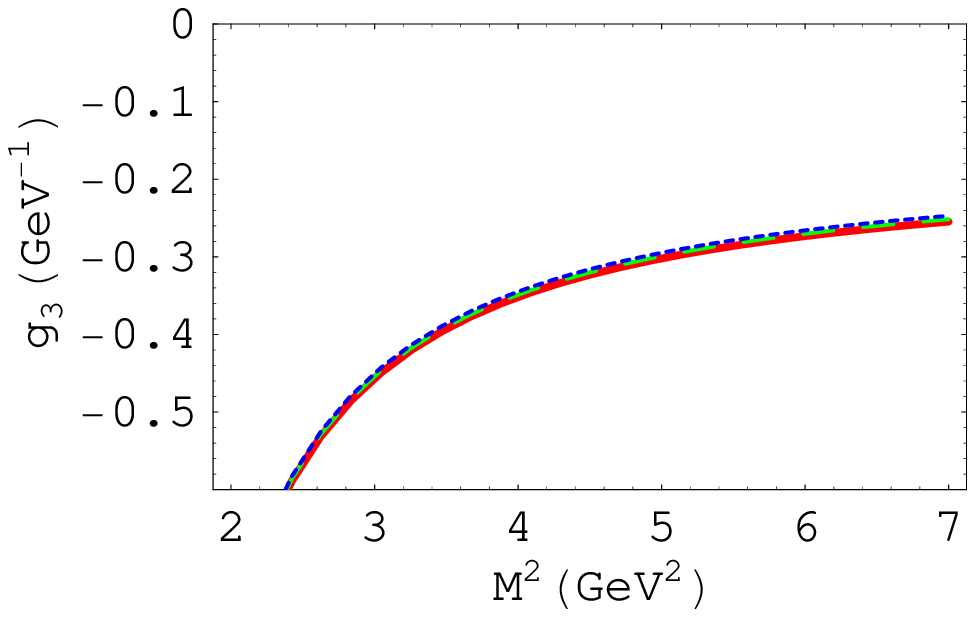, width=0.3\textwidth}\\
\end{tabular}
\end{center}
\caption{Sum rule results for the hadronic parameters in 
%eqs.(\ref{eq:ampDs0})-(\ref{eq:ampDs1Ds0}).
eqs.(2.1-4).
$M^2$ is the Borel parameter.}
\label{fig:results}
\end{figure}

The $D^*_{sJ}(2317) \to D_s^* \gamma$ and  $D_{sJ}(2460) \to D_s^{(*)} \gamma, \,  D_{sJ}^*(2317) \gamma$ decay amplitudes:
\bea
\langle \gamma(q,\lambda) D_s^*(p,\lambda^\prime)| D_{s0}(p+q)\rangle &=&  e d  \left[ (\varepsilon^* \cdot \tilde \eta^*)(p\cdot q)-(\varepsilon^* \cdot p)(\tilde \eta^* \cdot q) \right]   \label{eq:ampDs0} \\
\langle \gamma(q,\lambda) D_s(p)| D_{s1}^\prime(p+q,\lambda^{\prime\prime})\rangle &=&  e g_1  \left[ (\varepsilon^* \cdot  \eta)(p \cdot q)-(\varepsilon^* \cdot p)(\eta \cdot q) \right]   \label{eq:ampDs1a} \\
\langle \gamma(q,\lambda) D^*_s(p,\lambda^\prime)| D_{s1}^\prime(p+q,\lambda^{\prime \prime})\rangle &=& i \, e \, g_2 \,  \varepsilon_{\alpha \beta \sigma \tau}  \eta^\alpha \tilde \eta^{*\beta} \varepsilon^{*\sigma}   q^\tau   \label{eq:ampDs1b} \\
\langle \gamma(q,\lambda) D_{s0} (p)| D^\prime_{s1}(p+q,\lambda^{\prime \prime})\rangle &=& i \, e \,  g_3  \, \varepsilon_{\alpha \beta \sigma \tau}  \varepsilon^{*\alpha} \eta^\beta    p^\sigma q^\tau 
\label{eq:ampDs1Ds0}
\eea
($\varepsilon(\lambda)$ is  the photon 
 polarization vector and $\tilde \eta(\lambda^\prime)$,   
$\eta(\lambda^{\prime\prime} )$ the $D_s^*$  and $D_{s1}^\prime$ polarization vectors)
  involve the hadronic parameters $d, g_1, g_2$ and $g_3$. They can be determined
 by Light-Cone Sum rules \cite{altri,Colangelo:2000dp},
starting from the light-cone expansion of   correlation functions 
\be
F(p,q)=i \int d^4x \; e^{i p \cdot x} \langle \gamma(q,\lambda) | T[J^\dagger_A(x) J_B(0)] |0\rangle
\label{eq:corr-Ds0Ds*gamma}
\ee
of  quark-antiquark currents  $J_{A,B}$ having the same quantum number of the decaying and of the produced
charmed mesons,  and an
external photon state of momentum $q$ and helicity $\lambda$. 
%
%%%%%%%%%%%%%%%%%%%%%%%%%%%%%%%%%%%%%%%%%%%%%%%%
\begin{figure}[h]
 \begin{center}
\includegraphics[width=0.7\textwidth] {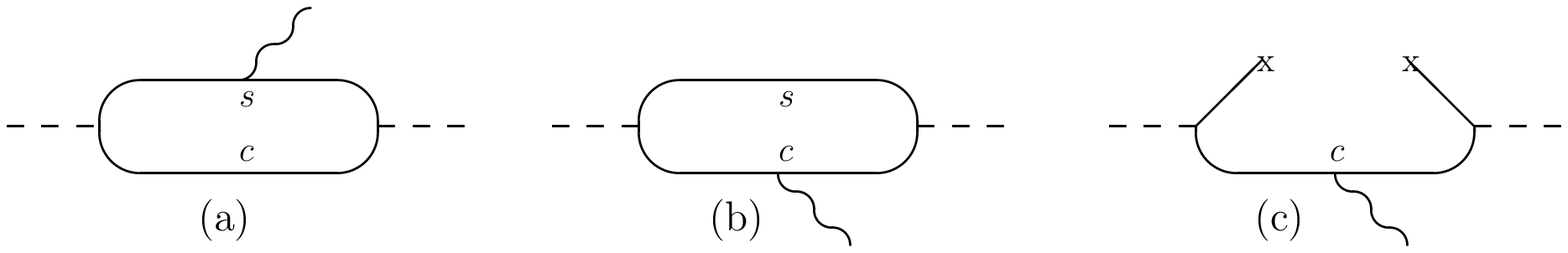}\\
\includegraphics[width=0.5\textwidth] {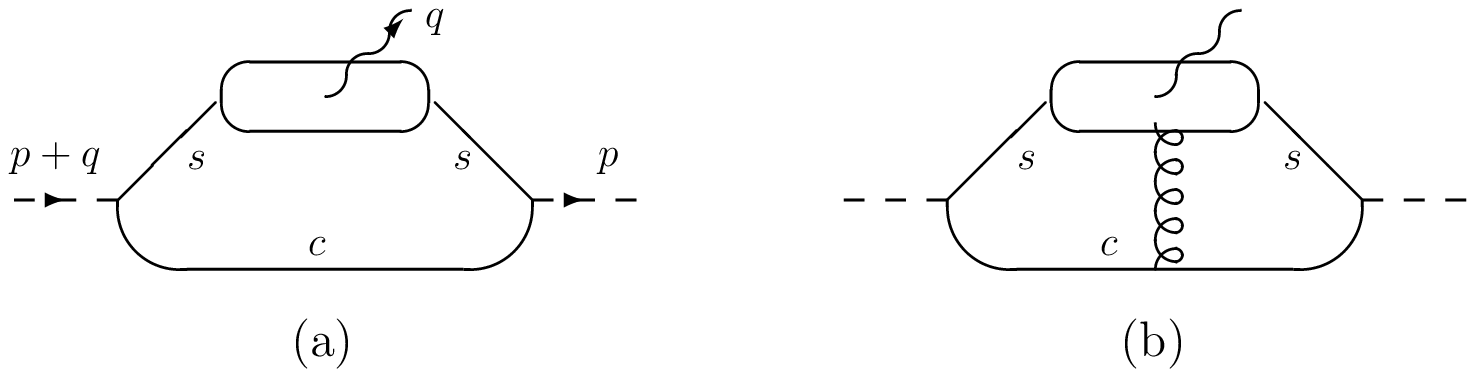}\\
\vspace*{0mm}
 \caption{ Diagrams describing the
 perturbative photon emission by the strange  and  charm  quark ( (a,b) in the first line)
 and  two and three-particle photon distribution amplitudes (second line);
 (c)  corresponds to the strange quark condensate contribution.}
  \label{fig:light-cone}
 \end{center}
\end{figure}
%%%%%%%%%%%%%%%%%%%%%%%%%%%%%%%%%%%%%%%%%%%%%%%%%
%
In this expansion  the perturbative  photon coupling to the strange and charm quarks is
considered together with  the contributions of the photon emission from the soft $s$ quark, 
expressed as photon matrix elements of increasing twist, fig.\ref{fig:light-cone} 
\cite{Ball:2002ps}. The correlation function can also be expressed in terms
of the contribution of the lowest-lying resonances,  the current-vacuum matrix elements of which 
are computed by the same method  \cite{Colangelo:1995ph}, 
 and a continuum of states which is treated invoking  global quark-hadron duality.
 A Borel transformation  introduces an external parameter $M^2$, 
  the hadronic quantities being
 independent of it. The results  depicted in fig. \ref{fig:results}  correspond to the rates in 
 Table \ref{predictions}.
%
%%%%%%%%%%%%%%%%%%%%%%%%%%%%%%%%%%%%%%%%%%%%
\begin{table}[h]
\caption{
Radiative decay widths (in keV) of $D^*_{sJ}(2317)$ and $D_{sJ}(2460)$ obtained by Light-Cone sum rules
(LCSR).   Vector Meson Dominance (VMD) and constituent quark model (QM) results
are also reported.} \label{predictions}
    \begin{center}
    \begin{tabular}{c c c c c c}
\hline Initial state & Final state & LCSR  \cite{Colangelo:2005hv}&VMD \cite{Colangelo:2004vu,colangelo1}
& QM \cite{Godfrey:2003kg}  & QM \cite{Bardeen} \\ \hline 
$D^*_{sJ}(2317)$&$D_{s}^{\ast }\gamma$&  4-6     & 0.85& 1.9 &  1.74\\ 
$D_{sJ}(2460)$   &$D_{s}\gamma$             & 19-29 & 3.3&  6.2 & 5.08 \\
                              &$D_{s}^*\gamma$          & 0.6-1.1 & 1.5&  5.5 & 4.66 \\
                               &\,\,\,\,\,$D^*_{sJ}(2317)\gamma$\,\,\,\,\,& 0.5-0.8  &  \ \hfill --- \hfill\ &0.012 & 2.74\\ \hline
\end{tabular}
\end{center}
\end{table}
%%%%%%%%%%%%%%%%%%%%%%%%%%%%%%%%%%%%%%%%%%%%

The main  result is that the rate of $D_{sJ}(2460) \to D_{s}\gamma$  is the largest one among the radiative  $D_{sJ}(2460)$ rates, and this is confirmed by 
experiment, see Table \ref{br}.
Since this result comes from a description of the mesons as  quark-antiquark states, we consider it as a quantitative argument in support of a $\bar c s$ interpretation of $D^*_{sJ}(2317)$ and  $D_{sJ}(2460)$.  The observation of all
the radiative decay modes with the predicted relative rates  would 
reinforce the conclusion. 
%
%%%%%%%%%%%%%%%%%%%%%%%%%%%%%%%%%%%%%%%%%%%%
\begin{table}[h]
\caption{Measurements and 90\% CL upper limits of  ratios of  $D^*_{sJ}(2317)$  and  $D_{sJ}(2460)$  decay widths.  } \label{br}
    \begin{center}
    \begin{tabular}{cccc}
\hline & Belle & BaBar   & CLEO  \\ \hline $\Gamma \left(
D^*_{sJ}(2317) \rightarrow D_{s}^{\ast }\gamma \right)/ \Gamma \left(
D^*_{sJ}(2317)\rightarrow D_{s}\pi ^{0}\right) $ & $<0.18$&  \ \hfill --- \hfill\   & $<0.059$ \\
$\Gamma \left(D_{sJ}(2460) \rightarrow
D_{s}\gamma \right) /\Gamma \left( D_{sJ}(2460)\rightarrow
D_{s}^{\ast }\pi ^{0}\right) $ & $0.45 \pm 0.09$ & $0.30 \pm 0.04$ & $<0.49$ \\
$\Gamma \left( D_{sJ}(2460)\rightarrow
D_{s}^{\ast }\gamma \right) /\Gamma \left( D_{sJ}(2460)\rightarrow D_{s}^{\ast }\pi ^{0}\right) $ & $<0.31$ &\ \hfill --- \hfill\ &$<0.16$ \\
$\Gamma \left( D_{sJ}(2460)\rightarrow
D^*_{sJ}(2317)\gamma \right)/ \Gamma \left( D_{sJ}(2460)\rightarrow
D_{s}^*\pi^0 \right) $  & \ \hfill --- \hfill\ & $ < 0.23$  & $ < 0.58$\\
\hline
\end{tabular}
\end{center}
\end{table}
%
%%%%%%%%%%%%%%%%%%%%%%%%%%%%%%%%%%%%%%%%%%%%
%
In order to quantitatively understand the  data in Table \ref{br} one should precisely know
the  widths of the isospin violating transitions
$D_{s0}\to D_s \pi^0$ and $D^\prime_{s1}\to D_s^* \pi^0$. In the description of    
these  transitions based on the mechanism of $\eta-\pi^0$ mixing 
\cite{wise,close,Godfrey:2003kg,colangelo1} the accurate determination of  the
strong   $D_{s0} D_s \eta$ and $D^\prime_{s0} D^*_s \eta$   couplings
for finite heavy quark mass and including $SU(3)$ corrections is required.   
\section{Narrow states in  the beauty sector}
The mass estrapolation from charm to beauty and the interpretation of 
$D^*_{sJ}(2317)$ and  $D_{sJ}(2460)$ as ordinary mesons allow  to predict the masses
of corresponding states in the beauty sector:
 $M(B_{s0})=5.71\pm 0.03 \,{\rm GeV}$ and $M(B^\prime_{s1})=5.77\pm 0.03 \,{\rm GeV}$
  \cite{colangelo2}.
 Therefore,
$B_{s0}$ and  $B^\prime_{s1}$ should be respectively below the $BK$ and $B^*K$ thresholds;
they should be observed as narrow peaks in   $B_s \pi^0$ and 
$B_s^* \pi^0$, $B^{(*)}_s\gamma$ invariant mass distributions, a possible observation  at 
 the Tevatron and the LHC.

\end{document}